\documentclass[twocolumn,aps,pra,showpacs,amsfonts,amssymb,amsmath,a4paper,flequ]{revtex4-1}
\usepackage{mathrsfs}
\usepackage{amsmath}
\usepackage{subfigure}

\usepackage[title,titletoc,header]{appendix}

\usepackage{graphicx}

\begin{document}
 \draft
 \title{ Particle-hole bound states of dipolar molecules in optical lattice}
 \author{Yi-Cai Zhang$^{1}$, Han-Ting Wang$^{1}$, Shun-Qing Shen$^{2}$, and Wu-Ming Liu$^{1}$}
 \address{$^{1}$Beijing National Laboratory for Condensed Matter
 Physics, Institute of Physics, Chinese Academy of Sciences, Beijing
 100190, China\nonumber\\
        $^{2}$Department of Physics and Centre of Theoretical and Computational Physics,
The University of Hong Kong, Pokfulam Road, Hong Kong, China}
\date{\today}

\begin{abstract}

We investigate the particle-hole pair excitations of dipolar molecules in optical
lattice, which can be described with an extended Bose-Hubbard model. For
strong enough dipole-dipole interaction, the particle-hole pair excitations can form bound states in one and two dimensions. With decreasing
dipole-dipole interaction, the energies of the bound states increase and
merge into the particle-hole continuous spectrum gradually. The existence regions, the energy
spectra and the wave functions of the bound states are carefully
studied and the symmetries of the bound states are analyzed with
group theory. For a given dipole-dipole interaction, the number of bound states
varies in momentum space and a number distribution of the bound states is
illustrated. We also discuss how to observe these bound states in future
experiments.

\end{abstract}

\maketitle

\section{Introduction}

Cold and ultracold molecules are attracting more and more attention due to
their broad applications in the fields of high-precision measurement,
quantum chemistry, quantum information and many-body physics \cite%
{molecule-review}. Recent years the experimental techniques for cold and
ultracold molecules have been developed greatly. The homonuclear molecular
Bose-Einstein condensation (BEC) was realized experimentally \cite%
{Jochim,Greiner} and the quantum state with exactly one molecule at each
site of an optical lattice was created with a Feshbach resonance and STIRAP
(stimulated Raman adiabatic passage) techniques \cite{Volz,Danzl1}.
Meanwhile the ultracold heteronuclear molecules were also produced \cite%
{Sage,Wang,Sawyer,Ni,Ospelkaus}.

The heteronuclear molecules, such as SrO, RbCs, or NaCs, prepared in their
electronic and vibrational ground states, have considerable permanent
electric dipole moment. 
The dipole-dipole interaction between molecules can be generated by the
external applied electric field \cite{molecule-review}. Moreover, the
magnitude of the dipole-dipole interaction can be controlled by the strength
of the electric field \cite{DeMille,buchler}. The tunable long range
dipole-dipole interaction may significantly modify the ground state and
collective excitations of trapped condensates. For example, in trapped
dipolar gases, it was shown theoretically \cite{Santos} that the mean-field
inter-particle interaction and, hence, the stability diagram are governed by
the trapping geometry. With increasing dipolar interaction, the ground state
of rotating atomic Bose gases undergoes a series of transitions between
vortex lattices of different symmetries: triangular, square, ``stripe", and
``bubble" phases \cite{Cooper}. In rapidly rotating Fermion gas, the
dipole-dipole interaction may even result in the fractional quantum
Hall-like states \cite{Baranov05}.

The particles with long-range interactions in optical lattice can be
described with extended Hubbard model \cite{goral,Menotti,Lin}. Compared with the regular Bose-Hubbard model with on-site interaction, the
extended Bose-Hubbard model has richer ground state phases, such as Mott
insulator, particle density wave, superfluidity or supersolid phase \cite%
{Bruder,Otterlo,Niyaz,Sengupta,Hassan,Iskin,PAI,Chen}. On the other hand,
the gapful particle and hole excitations in the insulating phase \cite%
{fisher,stoof,Kovrizhin} may bind together and form bound states due to the
long-range interactions. Although the excitons (holon-doublon pairs) in
one-dimensional fermionic Hubbard model have been extensively studied \cite%
{fermion-exciton}, the excitons in higher dimensions, and especially, in
bosonic Hubbard models are less studied.

In this paper, we study the particle-hole pair excitations of dipolar bosonic
molecules in optical lattice, especially, the possible bound states due to
the dipole-dipole interaction. The paper is organized as follows. In Sec.
\textbf{II} we introduce the extended Bose-Hubbard model to describe the
polarized bosonic molecules in optical lattice, and derive the eigen
equations to describe the single particle-hole pair excitation. In Sec.
\textbf{III}, the existence regions, the energies and the wave functions of
the particle-hole bound states are calculated in one and two
dimensions. The symmetries of the bound states are analyzed and possible
experimental observation of these bound states is also discussed. A summary
is presented in Sec. \textbf{IV}.

\section{The particle-hole bound states in the extended Bose-Hubbard model}

\begin{figure}[tbp]
\centerline{
\subfigure[]{\includegraphics[width=1.72in,height=2in]{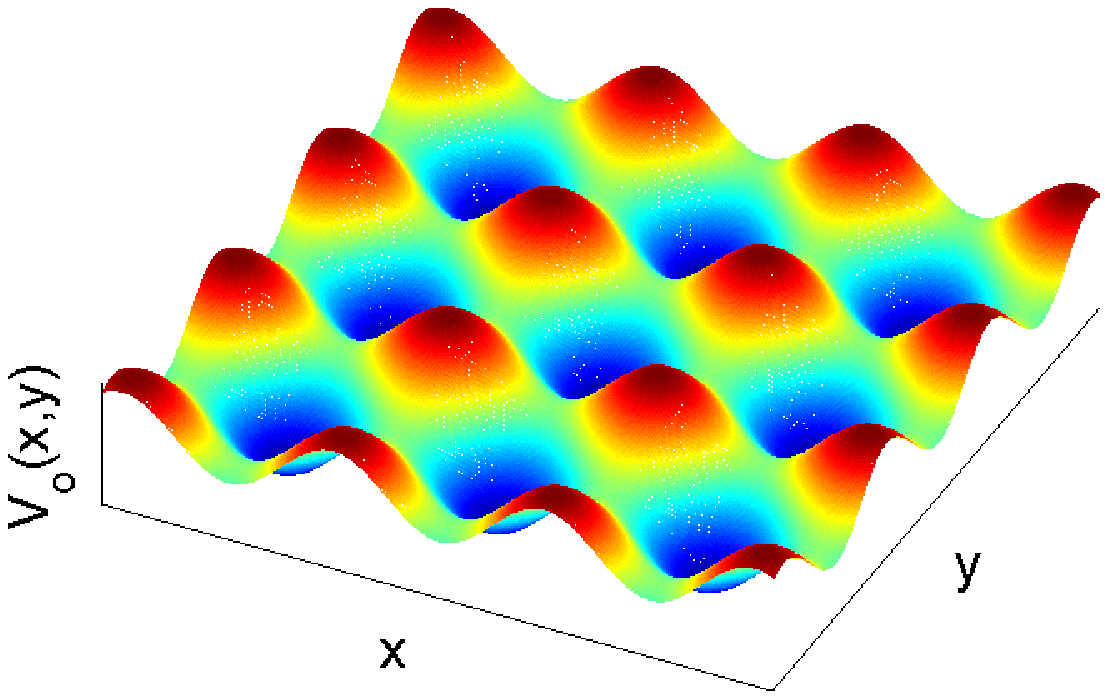}}
\subfigure[]{\includegraphics[width=1.72in,height=1.5in]{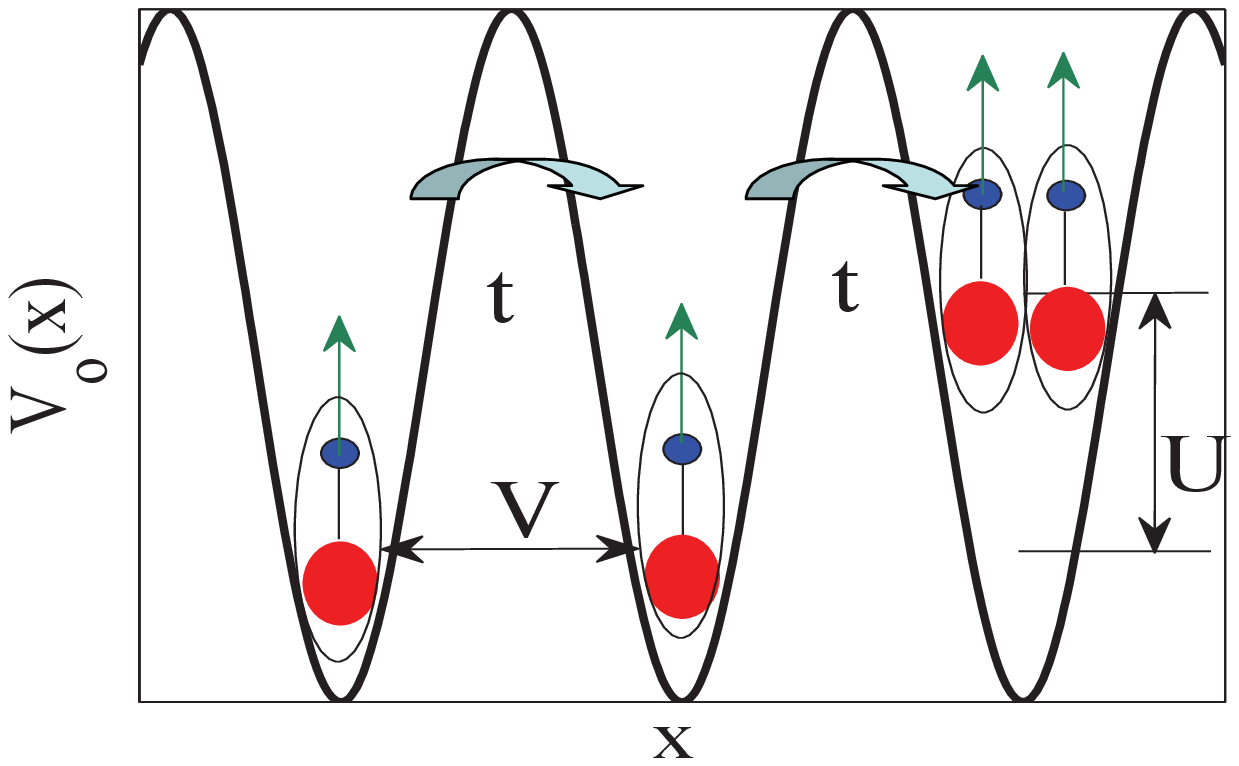}}}
\caption{Sketches of optical lattice and interactions of ultracold
heteronuclear molecules. (a) The landscape of optical lattice potential in
the x-y plane. The square optical lattice potential $%
V_{o}(x,y)=V_{opt}[sin^2(k_{x}x)+sin^2(k_{y}y)]$ can be created by counter
propagating far detuned laser beams, where $V_{opt}$ is the depth of lattice
and $k_{x(y)}$ is the laser wave vector along $x(y)$ direction.
(b) The interactions between dipolar molecules in the optical lattice (see
equation(1)). The arrows on the molecules indicate the polarization of the
electric dipole moments. }
\end{figure}

Considering only the nearest-neighbor interaction V to simulate the effect of dipole-dipole interaction, we write the
extended Bose-Hubbard model as,

\begin{align}
& H=H_{0}+H_{1},  \notag \\
& H_{0}=\frac{U}{2}\sum_{\vec{r}}n_{\vec{r}}(n_{\vec{r}}-1)+\frac{V}{2}\sum_{%
\vec{r},\vec{\sigma}}n_{\vec{r}}n_{\vec{r}+\vec{\sigma}},  \notag \\
& H_{1}={-t}\sum_{\vec{r},\vec{\sigma}}(b_{\vec{r}}^{\dagger }b_{\vec{r}+%
\vec{\sigma}}),
\end{align}%
where $n_{\vec{r}}=b_{\vec{r}}^{\dag }b_{\vec{r}}$ is the number operator at
site $\vec{r}$ with $b_{\vec{r}}(b_{\vec{r}}^{\dag })$ the annihilation
(creation) operator of particle. $\vec{\sigma}$ denotes the nearest neighbor
vectors of site $\vec{r}$. The on-site interaction $U$, the nearest neighbor
interaction $V$ and the hopping $t$ are expressed as
\begin{eqnarray}
U &=&g\int d\vec{x}|W(\vec{x})|^{4}, \notag \\
V &=&\int d\vec{x}d\vec{x}^{\prime }|W(\vec{x}-\vec{r})|^{2}V_{dd}(\vec{x}-%
\vec{x}^{\prime })|W(\vec{x}^{\prime }-(\vec{r}+\vec{\sigma}))|^{2}, \notag \\
t &=&-\int d\vec{x}W^{\ast }(\vec{x}-\vec{r})(-\frac{\hbar ^{2}}{2m}\nabla
^{2}+V_{0}(\vec{x}))W(\vec{x}-(\vec{r}+\vec{\sigma})),\notag \\
\end{eqnarray}
where $W(\vec{r})$ is Wannier function corresponding to the lowest energy
band and $V_{dd}(\vec{x})=\frac{C_{dd}}{4\pi }\frac{1-3cos^{2}\theta }{|\vec{%
x}|^{3}}$ is the dipole-dipole interaction of two particles separated with a
distance $|\vec{x}|$. For the electric dipole-dipole interaction, $C_{dd}=%
\frac{d^{2}}{\epsilon_0}$ with $d$ the electric dipole moment, $\epsilon_0$ is vacuum permittivity. We assume
that the polarization of dipole moment is along the $z$ axial
direction. In the following studies, we consider particles confined in
one-dimensional chain or two-dimensional square lattice and treat hopping
term $H_{1}$ as a perturbation.

In the atomic limit ($t=0$), $n_{\vec{r}}$ is a good quantum number and the
eigenstates of $H_{0}$ can be written as direct product of number states.
When the filling factor is one, $V\geq 0$ and $zV<U$, the Mott insulating
ground state can be written as $|11111111...\rangle $ with the ground state
energy $E_{0}={\frac{1}{2}}NzV$, where z is the coordination number and N
the number of lattice sites. The excited states contain one or more
particle-hole pairs. A single particle-hole pair state with a hole at $r$
and a particle at $r^{\prime }$ can be represented as $|r,r^{\prime }\rangle
=|110_{r}111112_{r^{\prime }}11...\rangle $. Taking into account the hopping
term $t$, the particle and hole will move in the lattice. Similar to the
two-magnon states in ferromagnetic system \cite{mattis}, the single
particle-hole pair state can be written as a linear combination of $%
|r,r^{\prime }\rangle $:
\begin{equation}
|\psi \rangle =\sum_{\vec{r},\vec{r}\prime }\phi _{\vec{r},\vec{r}\prime }|%
\vec{r},\vec{r}\prime \rangle ,
\end{equation}%
where $\phi _{\vec{r},\vec{r}^{\prime }}$
will be determined by solving the approximate eigen equation in single
particle-hole pair subspace
\begin{equation}
\langle \vec{r},\vec{r}\prime |H|\psi \rangle =E\langle \vec{r},\vec{r}%
\prime |\psi \rangle .
\end{equation}%

Calculating $H|\vec{r},\vec{r}^{\prime}\rangle$ and considering the boundary condition $\phi_{\vec{r%
},\vec{r}}=0$ (a particle and a hole do not share the same lattice site), we get

\begin{align}
&-\sum_{\vec \sigma}(t\phi_{\vec{r}+\vec \sigma,\vec{r}^{\prime}}+2t\phi_{%
\vec{r},\vec{r}^{\prime}+\vec \sigma}) =\varepsilon\phi_{\vec{r},\vec{r}%
^{\prime}}  \notag \\
&+\sum_{\vec \sigma}\delta_{\vec{r}^{\prime}-\vec{r},\vec \sigma}[V\phi_{%
\vec{r},\vec{r^{\prime}}} +(-t)\phi_{\vec{r}+\vec \sigma,\vec{r^{\prime}}%
}+(-2t)\phi_{\vec{r},\vec{r^{\prime}}-\vec \sigma}]  \notag \\
&-\delta_{\vec{r}^{\prime}-\vec{r},0}\sum_{\vec \sigma}(t\phi_{\vec{r}+\vec
\sigma,\vec{r}^{\prime}}+2t\phi_{\vec{r},\vec{r}^{\prime}+\vec \sigma}),
\end{align}
where $\delta_{\vec{r}^{\prime}-\vec{r},\sigma}$ is the Kronecker delta
function and  $\varepsilon=E-E_{0}-U$. The particle-hole excitation energy is $%
\omega=E-E_{0}=\varepsilon+U$.%

Owing to the translational invariance, it is convenient to apply a
transformation $\phi_{\vec{r},\vec{r}^{\prime}}=\frac{1}{\sqrt{N}}e^{i\vec{K}%
\cdot\vec{R}}\phi(\vec{\rho})$, where $\vec{R}=\frac{\vec{r}+\vec{r}^{\prime}%
}{2}$ and $\vec{\rho}=\vec{r}^{\prime}-\vec{r}$ denote the center-of-mass
and the relative coordinates respectively and $\vec{K}$ the total momentum
of the particle and hole.  At fixed $\vec K$, the eigen equation $(5)$
becomes

\begin{align}
& -\sum_{\vec{\sigma}}[te^{i\frac{\vec{K}\cdot \vec{\sigma}}{2}}\phi (\vec{%
\rho}-\vec{\sigma})+2te^{i\frac{\vec{K}\cdot \vec{\sigma}}{2}}\phi (\vec{\rho%
}+\vec{\sigma})]=  \notag \\
&\varepsilon \phi (\vec{\rho})+\sum_{\vec{\sigma}}\delta _{\vec{\rho},\vec{%
\sigma}}[V\phi (\vec{\rho})+(-te^{i\frac{\vec{K}\cdot \vec{\sigma}}{2}%
}-2te^{i\frac{-\vec{K}\cdot \vec{\sigma}}{2}})\phi (\vec{\rho}-\vec{\sigma})]
\notag \\
& -\delta _{\vec{\rho},0}\sum_{\vec{\sigma}}[te^{-i\frac{\vec{K}\cdot \vec{%
\sigma}}{2}}\phi (\vec{\sigma})+2te^{i\frac{\vec{K}\cdot \vec{\sigma}}{2}%
}\phi (\vec{\sigma})].  \label{Eq-5}
\end{align}%
\begin{figure}[tbp]
\begin{center}
\includegraphics[ scale=0.70 ]{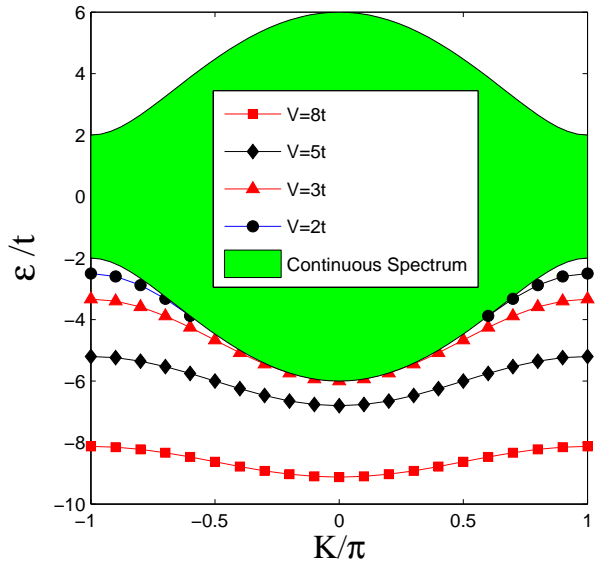}\\[0pt]
\end{center}
\caption{Energy spectra of the bound states of particle-hole pair in one dimension for
different nearest-neighbor interactions $V=2t$ (black circles), $V=3t$ (red
triangles), $V=5t$ (black diamonds) and $V=8t$ (red squares) respectively ($K=K_x$).
As comparison, the continuous spectrum of a particle-hole pair excitation is
also shown (green shaded region).}
\label{sw2d}
\end{figure}

Considering the finite range character of the nearest neighbor interaction $V$, we utilize
the Green's function approach to solve the eigen equation. To this end, we
introduce an effective Hamiltonian
\begin{equation}
H_{eff}=H_{ph,0}+V_{ph},  \label{Eq-6}
\end{equation}
where
\begin{align}
& H_{ph,0} & =& -t\sum_{\vec{\rho},\vec{\sigma}}e^{i\vec{K}\cdot \vec{\sigma}%
/2}|\vec{\rho}\rangle \langle \vec{\rho}-\vec{\sigma}|-2te^{i\vec{K}\cdot
\vec{\sigma}/2}|\vec{\rho}\rangle \langle \vec{\rho}+\vec{\sigma}|,  \notag
\\
& V_{ph} & =& \sum_{\vec{\sigma}}(2te^{-i\vec{K}\cdot \vec{\sigma}/2}+te^{i%
\vec{K}\cdot \vec{\sigma}/2})|\vec{\sigma}\rangle \langle 0|+h.c.  \notag \\
& & +& \sum_{\vec{\sigma}}(-V)|\vec{\sigma}\rangle \langle \vec{\sigma}|,\nonumber
\end{align}
to describe the motion of a particle around a hole, where $|\vec{\rho}\rangle$
 is basis set in relative coordinate spaces. The $\vec{K}$-dependent $%
H_{ph,0}$ describes the kinetic energy of the particle and $V_{ph}$ denotes
the interaction between particle and hole. With $|\phi \rangle =\sum_{\vec{%
\rho}}\phi (\vec{\rho})|\vec{\rho}\rangle $, it is easy to reproduce Eq. (%
\ref{Eq-5}) from the eigen equation $H_{eff}|\phi \rangle =\varepsilon|\phi
\rangle $. After a Fourier transformation, the kinetic energy is obtained as
\begin{eqnarray}
\varepsilon _{0}(\vec{K},\vec{q})&=&-t\sum_{\vec\sigma}e^{i(\vec{K}/2+\vec{q})\cdot{\vec\sigma}}
                                    -2t\sum_{\vec\sigma}e^{i(\vec{K}/2-\vec{q})\cdot{\vec\sigma}},\label{Eq-7}
\end{eqnarray}
where $\vec{q}$ is the relative momentum of particle and hole.
When a particle is adjacent to a hole, the energy offset $-V$ in
$V_{ph}$ may result in the formation of particle-hole bound states.
\begin{figure}[tbp]
\begin{center}
\includegraphics[ scale=0.7 ]{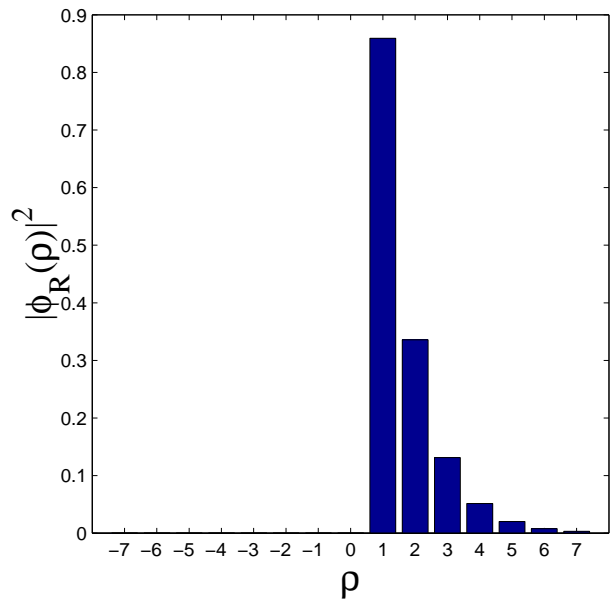}\\[0pt]
\end{center}
\caption{Probability density of a bound state in real space in the case of $%
V=8t$ and $K=0$. The probability density decreases exponentially with increase of distance. }
\label{sw2d}
\end{figure}

Introducing retarded Green functions $G_{0}=\lim_{\eta \to 0^{+}}
1/(\varepsilon-H_{ph,0}+i\eta)$ and $G= \lim_{\eta \to 0^{+}}
1/(\varepsilon-H_{eff}+i\eta)$, we could calculate $G$ through the
Lippmann-Schwinger equation $G=G_{0}+G_{0}V_{ph}G$. In the real
space, the Green's function $G$ has a general form

\begin{equation*}
\langle \vec{\rho ^{\prime }}|G|\vec{\rho}\rangle =\lim_{\eta \to 0^{+}}\sum_{m}\frac{\phi _{m}(%
\vec{\rho ^{\prime }})\phi _{m}^{\ast
}(\vec{\rho})}{\varepsilon-E_{m}+i\eta}=G(\vec{\rho ^{\prime
}},\vec{\rho},\varepsilon),
\end{equation*}%
where the contributions of the continuous spectra are neglected and $E_{m}$%
's are the eigenvalues of the bound states with $\phi _{m}$'s the
corresponding eigenfunctions \cite{Economou}. We can determine the
eigenvalues and eigenfunctions by analyzing the poles and residues of the
obtained Green's function.

\begin{figure}[tbp]
\begin{center}
\includegraphics[ scale=0.65 ]{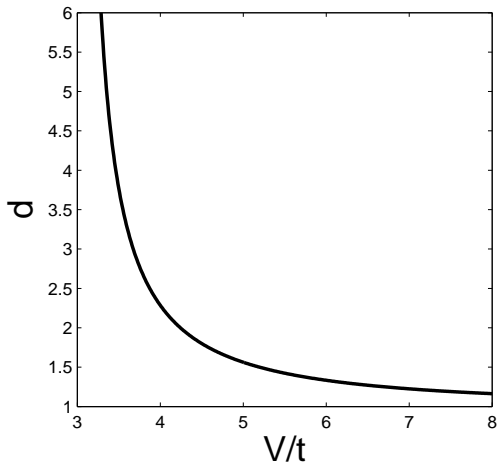}\\[0pt]
\end{center}
\caption{The mean size $d$ of bound states as a function of $V$ at K=0 in one
dimension. When the nearest-neighbor interaction $V$ approaches the critical
value $V=3t$, the mean size of the bound state becomes divergent
and the bound state energies merge into the continuous spectrum (see the red
triangles in Fig.2), indicating the disintegration of the bound states.}
\label{sw2d}
\end{figure}

\begin{figure}
\centering
\begin{minipage}[c]{0.25\textwidth}
    \centering
    \includegraphics[width=2in]{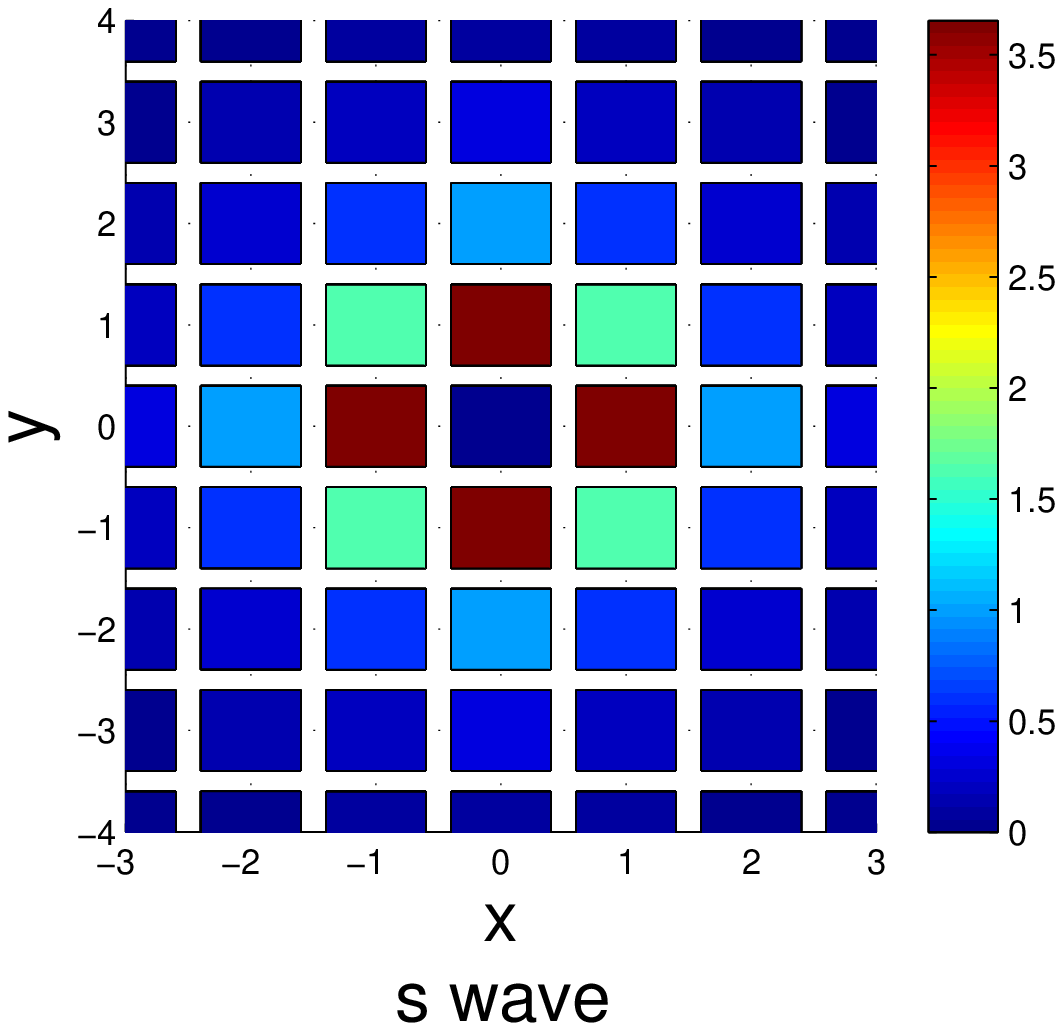}
  \end{minipage}
\begin{minipage}[c]{0.25\textwidth}
    \centering
    \includegraphics[width=2in]{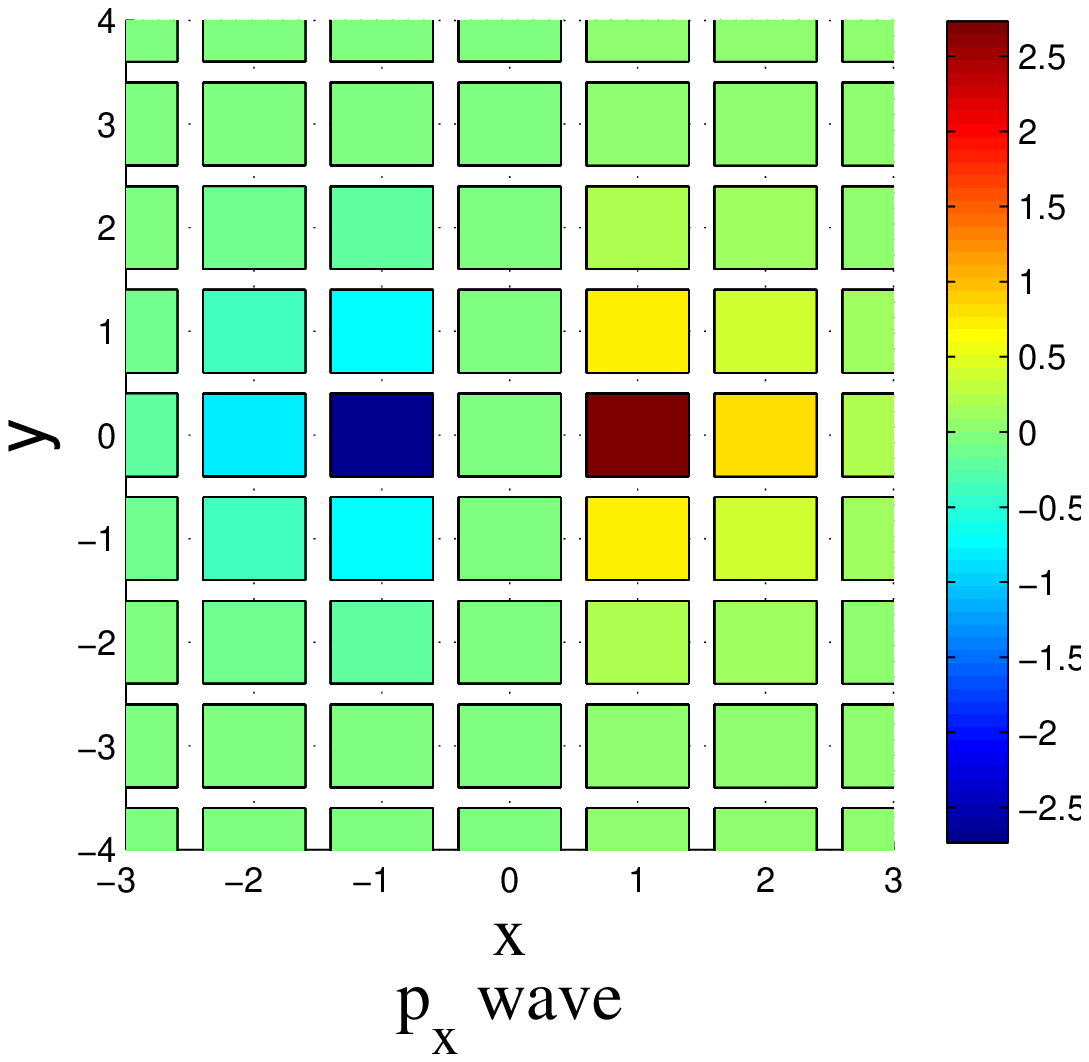}
  \end{minipage}
\begin{minipage}[c]{0.25\textwidth}
    \centering
    \includegraphics[width=2in]{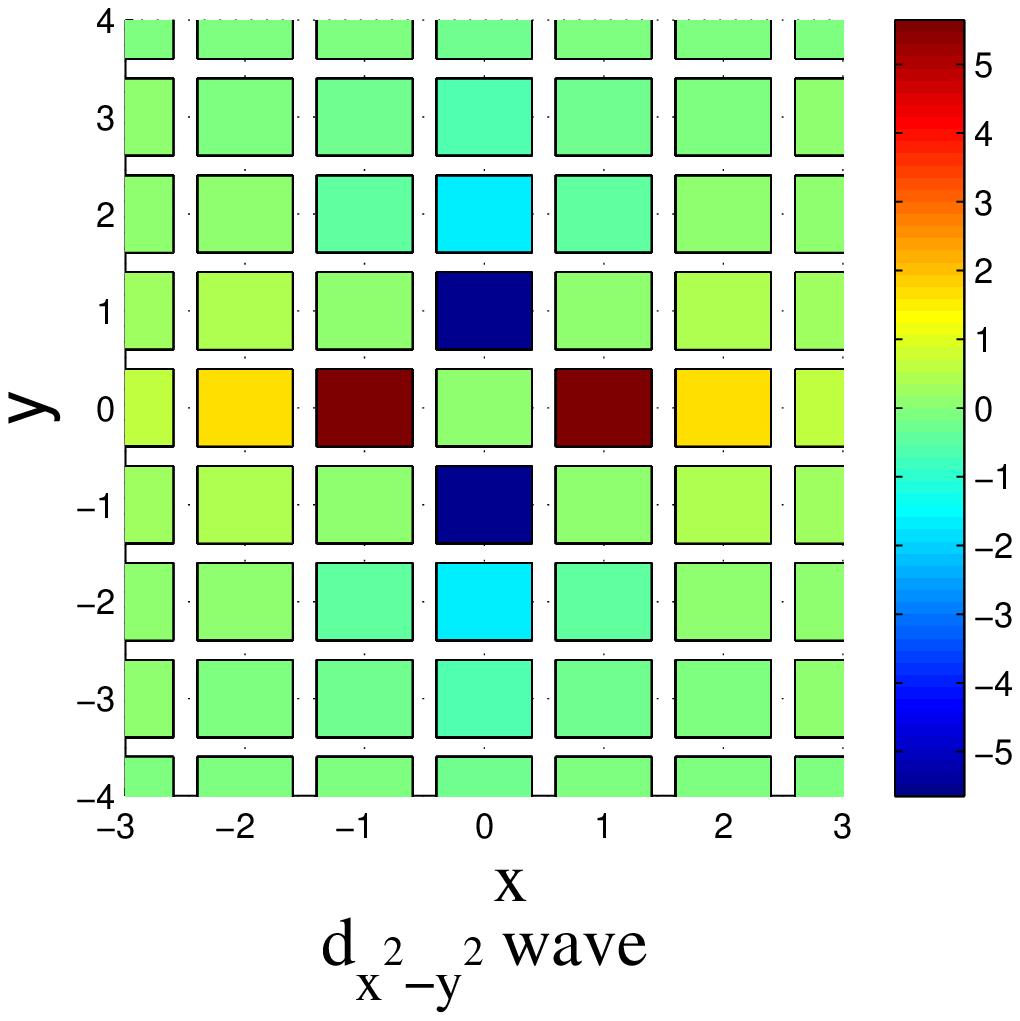}
  \end{minipage}
\caption{The wave functions of bound states in two dimensional case at $\vec{K}%
=(0,0)$ and $V=12t$ (non-normalized). $s$-, $p_x$- and $d_{x^2-y^2}$-wave
symmetries can be observed respectively.}
\label{sw2d}
\end{figure}

From Eq. (\ref{Eq-6}), we have
\begin{equation*}
\langle \vec{\rho ^{\prime }}|G_{0}|\vec{\rho}\rangle =\lim_{\eta \to 0^{+}}\frac{1}{(2\pi )^{d}}%
\int_{-\pi }^{\pi }\frac{d^{d}qe^{i\vec{q}\cdot (\vec{\rho ^{\prime }}-\vec{%
\rho})}}{\varepsilon-\varepsilon _{0}(\vec{K},\vec{q})+i\eta},
\end{equation*}%
and
\begin{eqnarray}
\langle \vec{\rho ^{\prime }}|G|\vec{\rho}\rangle  &=&\langle \vec{\rho
^{\prime }}|G_{0}|\vec{\rho}\rangle   \notag \\
&&+\sum_{\vec{\rho _{1}},\vec{\rho _{2}}}\langle \vec{\rho ^{\prime }}|G_{0}|%
\vec{\rho _{1}}\rangle \langle \vec{\rho _{1}}|V_{ph}|\vec{\rho _{2}}\rangle
\langle \vec{\rho _{2}}|G|\vec{\rho}\rangle .  \label{Eq-9}
\end{eqnarray}

For specific $\vec{\rho \prime }=0$ or $\sigma $, Eq. (\ref{Eq-9}) can be
reduced to a set of simultaneous linear equations. Green's functions $%
\langle \vec{\rho ^{\prime }}|G|\vec{\rho}\rangle $ with $\vec{\rho \prime }%
=0$ or $\sigma $ can thus be obtained exactly. The residues of  $\langle 0|G|\vec{\rho}%
\rangle $ is always vanishing for non-vanishing bound states energies $\varepsilon $ in our calculations,
which is consistent with the boundary condition $\phi (0)=0$.
 We present our results in the next section.

\section{bound states in one and two dimensions}

\subsection{Bound states in one dimension}

In the Mott insulating phase, the excitation energy of a single
particle/hole was calculated as
\begin{eqnarray}
\omega _{p,h} &=&\pm \lbrack -\frac{\varepsilon _{0}}{2}+U(n_{0}-\frac{1}{2}%
)+2dVn_{0}-\mu ]  \notag \\
&&+[(\frac{\varepsilon_{0}}{2})^{2}+\varepsilon _{0}U(n_{0}+\frac{1}{2})+\frac{%
U^{2}}{4}]^{\frac{1}{2}}
\end{eqnarray}
with the dynamical Gutzwiller approach \cite{Kovrizhin}. Up to the first
order of $\frac{t}{U}$, the single particle-hole pair excitation energy is
\begin{eqnarray}
\omega=\omega _{p}(\vec{k}_{p})+\omega _{h}(\vec{k}_{h})
&=&U+\varepsilon _{0}(\vec{K},\vec{q})
\end{eqnarray}
with the total momentum $\vec{K}=\vec{k}_{p}+\vec{k}_{h}$ and relative
momentum $\vec{q}=(\vec{k}_{h}-\vec{k}_{p})/2$. For the extended
Bose-Hubbard model, however, a particle-hole pair may form a bound state due
to the interaction (See Eq. (\ref{Eq-6})). In the following calculations, we
take the lattice constant as $1$ and use $t>0$ as the unit of energy.
Conditions of $t/U\ll 1$ and $zV<U$ are assumed to make sure that the ground
state of the system is the deep Mott insulating phase.

In one dimensional case, $\varepsilon
_{0}(\vec{K},\vec{q})=-2t_{\hat{x}}\cos(q+\theta_x)$ with $t_{\hat{x}}\!\!\!\!=\!\!t\sqrt{%
9cos^{2}(\frac{K_{x}}{2})+sin^{2}(\frac{K_{x}}{2})}$ and $\theta
_{\hat{x}}=arctan(\frac{1}{3}tan(\frac{K_{x}}{2}))$. The top and
bottom boundaries of particle-hole continuous spectrum are
\begin{equation}
\Omega _{1}^{t(b)}(K)=\pm 2t_{\hat{x}},
\end{equation}
(see Fig.2). Compared with particle-hole pairs in fermion Hubbard model with half filling \cite{Barford}, the minimum of band width of the continuous spectrum is non-zero at $K=\pm\pi$. This is because that there are no particle-hole symmetries in Bose-Hubbard model.

 For a specific nearest-neighbor interaction $V$ and
total momentum $K$, we search for bound state solutions outside the
continuum. In the case of $\varepsilon<-2t_{\hat{x}}$, the free
Green function is calculated as \cite{Nygaard}
 \begin{eqnarray}
\langle \vec{\rho ^{\prime }}|G_{0}|\vec{\rho}\rangle
  &=& \lim_{\eta \to 0^{+}}\frac{1}{(2\pi )^{d}}
     \int_{-\pi }^{\pi }\frac{d^{d}qe^{i\vec{q}\cdot (\vec{\rho ^{\prime }}-\vec{%
      \rho})}}{\varepsilon-\varepsilon _{0}(\vec{K},\vec{q})+i\eta}\notag \\
&=&\alpha\beta^{|\rho-\rho'|}\gamma^{\rho-\rho'} ,
\end{eqnarray}%
where $\alpha=-\frac{1}{\sqrt{\varepsilon^{2}-4t_{\hat{x}}^{2}}}$,
$\gamma=e^{i\theta_{\hat{x}}}$ and $\beta=e^{-\kappa}$ with
$\kappa=arccosh(\frac{|\varepsilon|}{2t_{\hat{x}}})$.

\begin{figure}[tbp]
\centerline{
\subfigure[]{\includegraphics[width=6.5cm,height=6.5cm]{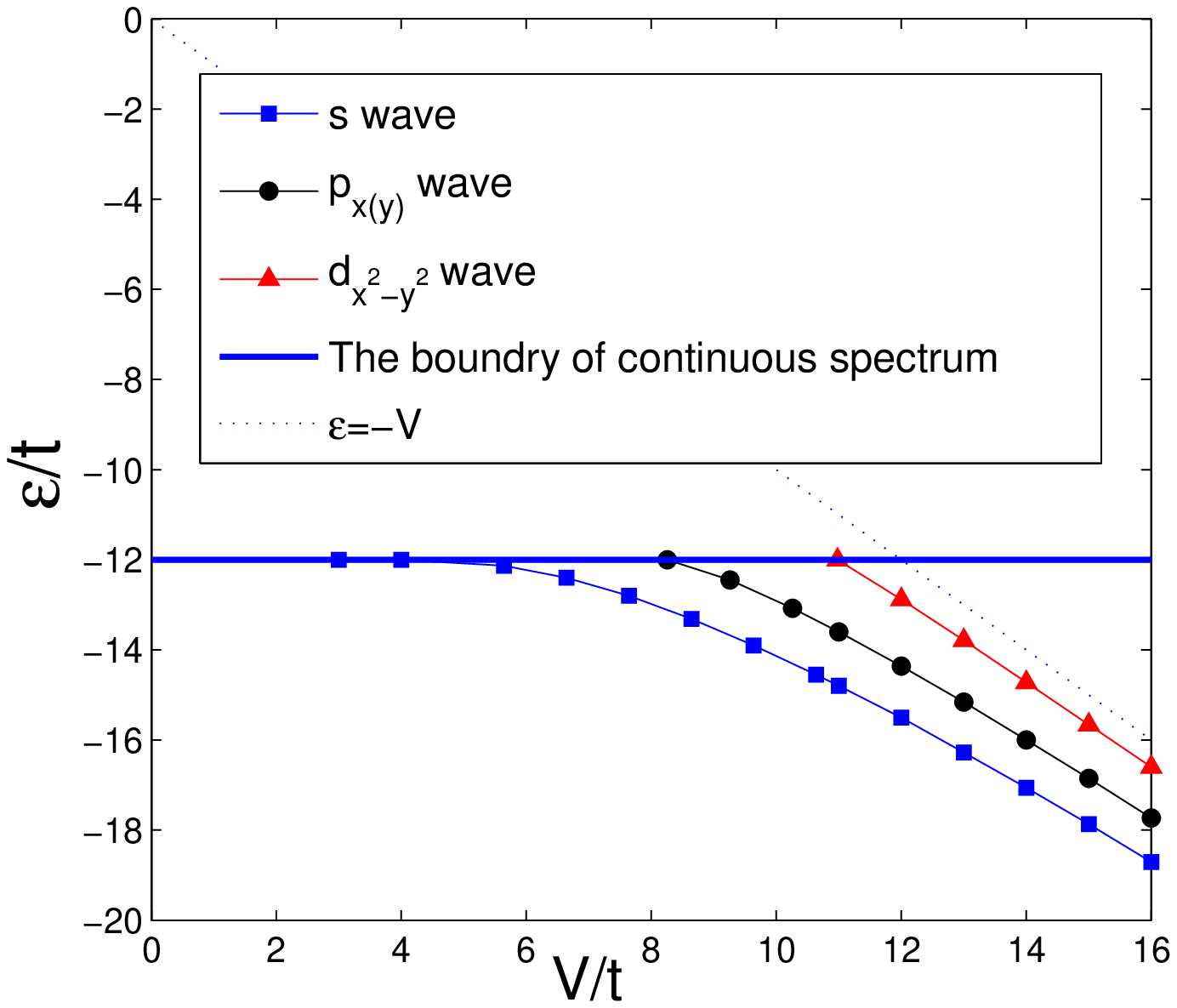}}}
\caption{Variations of the energies of different bound states in two
dimension at $\vec K=(0,0)$. The energies of different bound states increase
with decreasing $V$ and cross the particle-hole continuum ($-12t$ at $\vec
K=(0,0)$) at different $V$.}
\end{figure}

With $\rho ^{\prime }=0$, $\pm 1$, Eq. (\ref{Eq-9}) is reduced to
$3$ linear equations. The Green functions of
$\langle-1|G|\rho\rangle$, $\langle 0|G|\rho\rangle$ and $\langle
1|G|\rho\rangle$ can be exactly obtained as

\begin{align}
&\langle-1|G|\rho\rangle=\frac{\theta(-\rho)(\beta/\gamma)^{-(\rho+1)}}{-\frac{1}{\beta}(t_{\hat{x}}-V\beta)}&\rho\neq0 \notag\\
&\langle0|G|\rho\rangle=0,\notag\\
&\langle1|G|\rho\rangle=\frac{\theta(\rho)(\beta\gamma)^{\rho-1}}{-\frac{1}{\beta}(t_{\hat{x}}-V\beta)}&\rho\neq0,
\end{align}
where $\theta(\rho)$ is Heaviside step function.
 From these Green functions, we obtain two degenerate bound states corresponding to the particle on the left (right) of the
hole respectively. The bound state energy is
\begin{eqnarray}
\varepsilon=-V-\frac{t_{\hat{x}}^{2}}{V}.
\end{eqnarray}
Accordingly, $\beta$ is simplified as
$\beta=\frac{t_{\hat{x}}}{V}<1$ and the condition for the existence
of the bound state is obtained as
\begin{eqnarray}
  V\!\!>\!\!t\sqrt{9cos^{2}(\frac{K_{x}}{2})+sin^{2}(\frac{K_{x}}{2})}.
\end{eqnarray}
When $0 \leq V \leq t$, the interaction between particle and hole is
too weak to bind them together and no bound state is found. When
$t<V<3t$, two degenerate bound states exist in the regions of
$2arccos\sqrt{\frac{V^2-t^{2}}{8t^{2}}}<|K_x| \leq \pi$.  The region
get larger with the increase of $V$. When $V \geq 3t$, bound states
may be found in the whole first Brillouin zone. In Fig.2, we show
the spectra of the bound states for the interaction $V=2t$, $3t$,
$5t$ and $8t,$ respectively. The energies of the bound states
decrease with increasing $V$. At $V=2t$, the calculated existence
intervals are $0.58\pi<|K_x|\leq \pi$.

From the residues of the Green function, we can also extract the
bound state wave functions. For example, the bound state with a
particle on the right of a hole is written as£»
\begin{align}
\phi_R(\rho)&=C(\frac{t_{\hat{x}}}{V})^{(\rho-1)}e^{-i\theta_{\hat{x}}(\rho-1)}£¬&&\rho\geq1\notag\\
\phi_R(\rho)&=0£¬&&\rho<1,
\end{align}
where $C=\sqrt{1-\frac{t_{\hat{x}}^{2}}{V^2}}$ is the normalized
constant. In Fig.3, we show the probability distribution of the wave
function for $K=0$ and $V=8t$.

The mean size of the bound state is calculated as
$d=\langle\phi_{R}(\rho)|\rho|\phi_{R}(\rho)\rangle=\frac{V^2}{V^2-t_{\hat{x}}^{2}}$.
In Fig.4, we show the mean size of the bound state as a function of
the nearest-neighbor interaction $V$ at $K_x=0$. The stronger is the
interaction, the smaller is the size and the closer do the particle
and hole bind.

\begin{figure}[tbp]
\begin{center}
\includegraphics[ scale=0.60]{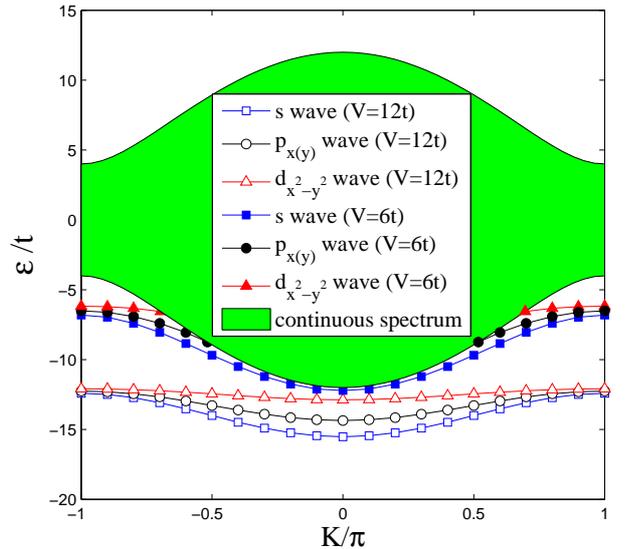}\\[0pt]
\end{center}
\caption{Dispersion relations of bound states along the line of $K_x=K_y=K$
for $V=6t$ and $V=12t$. As comparison, the particle-hole continuum is also
shown (green shaded region).}
\label{sw2d}
\end{figure}

\subsection{Bound states in two dimension}

Before presenting the numerical results, we discuss the symmetries of the
effective Hamiltonian $H_{eff}$ in Eq. (\ref{Eq-6}) in details. After a
gauge transformation
\begin{equation}
|\vec{\rho} \rangle ^{\prime }=|x,y\rangle ^{\prime }=e^{-i(x\theta _{\hat{x}}+y\theta
_{\hat{y}}})|x,y\rangle =e^{-i(x\theta _{\hat{x}}+y\theta _{\hat{y}}})|\vec{\rho} \rangle
\end{equation}
with $\theta _{\hat{x}}\!\!=\!\!arctan(\frac{1}{3}tan(\frac{K_{x}}{2}))$ and $\theta
_{\hat{y}}\!\!=\!\!arctan(\frac{1}{3}tan(\frac{K_{y}}{2}))$, we could remove the phase
factors in the Hamiltonian and get
\begin{eqnarray}
H_{eff}^{\prime } &=&H_{ph,0}^{\prime }+V_{ph}^{\prime }, \\
H_{ph,0}^{\prime } &=&-t_{\hat{x}}\sum_{\vec{\rho},\vec{\sigma}_{x}}|\vec{\rho}%
\rangle ^{\prime }\langle \vec{\rho}-\vec{\sigma}_{x}|^{\prime }-t_{\hat{y}}\sum_{%
\vec{\rho},\vec{\sigma}_{y}}|\vec{\rho}\rangle ^{\prime }\langle \vec{\rho}-%
\vec{\sigma}_{y}|^{\prime },  \notag \\
V_{ph}^{\prime } &=&-V\sum_{\vec{\sigma}}|\vec{\sigma}\rangle ^{\prime
}\langle \vec{\sigma}|^{\prime }+t_{\hat{x}}\sum_{\vec{\sigma}_{x}}|\vec{\sigma}%
_{x}\rangle ^{\prime }\langle 0|^{\prime }+h.c.  \notag \\
&&+t_{\hat{y}}\sum_{\vec{\sigma}_{y}}|\vec{\sigma}_{y}\rangle ^{\prime }\langle
0|^{\prime }+h.c.,  \notag
\end{eqnarray}%
where $\vec{\sigma}_{x}$ and $\vec{\sigma}_{y}$ denote the nearest
neighbors along the $x$ and $y$ directions, respectively, and $t_{\hat{x}}\!\!\!\!=\!\!t\sqrt{%
9cos^{2}(\frac{K_{x}}{2})+sin^{2}(\frac{K_{x}}{2})}$ and $t_{\hat{y}}\!\!=t\sqrt{%
9cos^{2}(\frac{K_{y}}{2})+sin^{2}(\frac{K_{y}}{2})}$ are effective hoppings
along the $x$ and $y$ directions.

When $K_{x}=\pm K_{y}$, $t_{\hat{x}}=t_{\hat{y}}$, $H_{eff}^{\prime }$ has symmetries
of $D_{4}$ group. According to the irreducible representations of this
group, the bound states can be classified and labeled with $s$, $p_{x(y)}$
and $d_{x^{2}-y^{2}}$ wave, respectively. Among them, the $s$ wave belongs
to an identical representation $A_{1}$ of $D_{4}$ group, the degenerate $%
p_{x}$ and $p_{y}$ waves form a two dimensional irreducible representation $E
$ of $D_{4}$ group, and the $d_{x^{2}-y^{2}}$ wave belongs to an irreducible
representation $B_{1}$ of $D_{4}$ group.

When $K_x\neq\pm K_y$, $t_{\hat{x}}\neq t_{\hat{y}}$, the symmetry reduces to $D_2$,
a subgroup of $D_{4}$. For simplicity, we still label the four bound states
with $s$, $p_{x(y)}$ and $d_{x^2-y^2}$. Differently, here all the $s$ and $d$
wave belong to identical representations $A_1$ of $D_{2}$ group, while $p_{x}
$($p_{y}$) wave belongs to an irreducible representation $B_{3}$($B_{2}$) of
$D_{2}$ group. The degeneracy of $p_x$ and $p_y$ waves is lifted.

With $\rho^{\prime}=(0,0)$, $(\pm 1,\pm 1)$, Eq. $(9)$ reduces to
five linear equations in two dimensional case. The free Green
function $G_{0}(\vec{\rho},\vec{\rho}')$ can be expressed with
elliptic integrals \cite{Economou,WORTIS}. Although no brief
solutions of $<\rho|G|\rho^{\prime}>$ could be found, we may
factorize the particle-hole bound state equations and analyze the
existence conditions for the bound states along the symmetric lines
of $K_{x}=\pm K_{y}=K$ \cite{WORTIS}. Considering the asymptotic
behaviours of elliptic integrals, we get the thresholds of the
interaction $V$ as follows: $V_{cr,s}(K)=t_{\hat{x}}$ for s wave,
$V_{cr,p}(K)=\frac{2\pi}{2\pi-4}t_{\hat{x}}$ for $p_{x(y)}$ wave and
$V_{cr,d}(K)=\frac{2\pi}{8-2\pi}t_{\hat{x}}$ for $d_{x^{2}-y^{2}}$
wave. From these thresholds, the existence region of every bound
state is obtained respectively. For $s$ wave, we have
\begin{eqnarray}
2arccos\sqrt{\frac{V^2-t^{2}}{8t^{2}}}\leq |K|\leq \pi, V\in( t,3t].
\nonumber
\end{eqnarray}
When $V>3t$, the $s$ wave bound state may be found in the whole
Brillouin zone. The existence region of $p$ wave bound states is
\begin{eqnarray}
2arccos\sqrt{\frac{(\frac{\pi-2}{\pi}V)^2-t^{2}}{8t^{2}}}\leq
|K|\leq \pi, V\in (\frac{\pi}{\pi-2}t, \frac{3\pi}{\pi-2}t].
\nonumber
\end{eqnarray}
When $V>\frac{3\pi}{\pi-2}t$, the existence region extends to the
whole Brillouin zone. For $d$ wave, existence region is expressed as
\begin{eqnarray}
2arccos\sqrt{\frac{(\frac{4-\pi}{\pi}V)^2-t^{2}}{8t^{2}}}\leq
|K|\leq \pi, V\in (\frac{\pi}{4-\pi}t,\frac{3\pi}{4-\pi}t].\nonumber
\end{eqnarray}
When $V>\frac{3\pi}{4-\pi}t$, the d wave existence interval is the
whole Brillouin zone.

Now we present the numerical results in two dimensions. At $\vec{K}=(0,0)$,
four bound states could be found when $V\geq V_{cr,d}(0)=10.97t$. The wave
functions with $s-$, $p_{x}-$ and $d_{x^{2}-y^{2}}-$wave symmetry for $V=12t$
are shown in Fig. 5. With the decrease of $V$, the highest $d_{x^{2}-y2}$%
-wave, the degenerate $p_{x(y)}$-waves and the lowest $s$-wave merge into
the continuous spectrum one by one. Finally, all the bound states disappear
when $V\leq V_{cr,s}(0)=3t$. In Fig. 6, we illustrate the variations of the bound
state energies with the changes of $V$.

We then search for bound state solutions along the line of
$K_{x}=K_{y}=K$. As shown in Fig. 7, there are four bound states for
all $K\in \lbrack -\pi ,\pi ]$ at $V=12t$. At $V=6t$, $s$-wave
exists in the whole region, while $p_{x(y)}$- and
$d_{x^{2}-y^{2}}$-wave states appear at $0.52\pi<|K|\leq \pi$ and
$0.69\pi<|K|\leq \pi$ respectively. Decreasing $V$ further, we find
that the regions of the bound states shrink, and all the bound
states disappear when $V<V_{cr,s}(\pm\pi)=1t$. Similar results are
obtained along the line of $K_{x}=-K_{y}=K$.

Away from the lines of $K_{x}=\pm K_{y}$, the degeneracy of $p_{x}$ and $%
p_{y}$ bound states is lifted, as mentioned before. At a given $V$,
different bound states have different existence regions in $\vec{K}$ space.
Consequently, the number of bound states vary in the $(K_{x},K_{y})$ space.
As an example, we show the number distribution of the bound states for $V=6t$
in Fig. 8.

\begin{figure}[tbp]
\includegraphics[width=5.5cm,height=5.5cm]{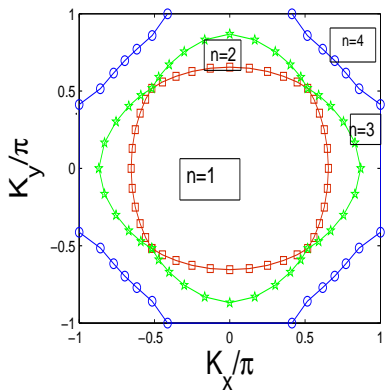}\newline
\caption{The number distribution of bound states in the first Brillouin zone
at $V=6t$. As shown, there are more bound stats at the corners than that at the center in the first Brillouin zone.}
\end{figure}

\subsection{Discussion of observations of the bound states}

Inelastic light scattering directly measures the dynamical structure factor $%
S(\vec q, \omega)$, the Fourier transformation of density correlations.
Bragg spectroscopy has been proposed to detect quantum phases in optical lattice and successfully applied to measure the
excitation spectra (phonons of BEC), the composition of the excitations and the Higgs-type
amplitude mode in the superfluid condensate, as well as the particle-hole
excitation energies in the Mott-insulator state \cite%
{Ye,kurn,Oosten,Rey,Ernst,Bissbort,Clement}. When this technique is utilized in the
study of ultra-cold polar molecules in the optical lattice, we may expect
extra resonance peaks corresponding to the particle-hole bound states lying
outside the particle-hole continuum.

Compared with the traditional solid state counterparts, the deep Mott
insulating state with $t/U\ll 1$ and $zV<U$ can be realized by tuning the
interaction parameters $t$, $U$ and $V$ in ultra-cold dipolar molecules in
the optical lattices. In deep optical lattice the hopping is approximately
evaluated as $t=(4/\sqrt{\pi }%
)E_{r}(V_{opt}/E_{r})^{3/4}exp[-2(V_{opt}/E_{r})^{1/2}]$, with $%
V_{opt}=18\sim 20E_{r}$ the optical lattice depth and $E_{r}$ the lattice
recoil energy \cite{Bloch}. Under the condition of $\frac{as}{a_{\bot }}\ll 1
$, where $a_{s}$ is the $s$-wave scattering length and $a_{\bot }=\sqrt{%
\frac{\hbar }{m\omega _{\bot }}}$ with $\omega _{\bot }$ transverse trapping
frequency in one dimension or trapping frequency of the $z$ direction in two
dimensions, the on-site interaction is estimated as $U_{1D}=\sqrt{\frac{2}{%
\pi }}\hbar \omega _{\bot }a_{s}/l_{0}$ and $U_{2D}=\frac{\sqrt{8\pi }\hbar
^{2}a_{s}}{ma_{z}(2\pi )l_{0}^{2}}$ with $l_{0}=(\frac{E_{r}}{V_{opt}})^{%
\frac{1}{4}}\frac{a}{\pi }$ and $a$ the lattice constant \cite{Jaksch,Bloch}.

Taking Bose molecule $\mathrm{{^{23}Na^{7}Li}}$ for example, we may estimate
the typical values of $t$, $U$ and $V$ for observing the particle-hole bound
state. The molecule prepared in the ground state has a permanent electric
moment $d=0.58D$ \cite{Mabrouk}. With a lattice constant $a\sim 0.5\mu m$, $%
a_{s}/a\sim 0.01$ and the transversal tapping frequency $\omega _{\bot }\sim
2\pi \times 10^{4}\mathrm{Hz}$, we have $t\sim 1nK$, $U_{1D}\sim 25nK$ and $%
U_{2D}\sim 50nK$. The nearest-neighbor interaction $V$ can be tuned as $%
V\sim 10nK$ with an applied electric field.

\section{Summary}

In summary, we have investigated bound states of particle-hole pair resulting from the
dipole-dipole interaction between polar molecules in the optical lattice.
For a large enough dipole-dipole interaction, two degenerate bound states,
which correspond to a particle on the left and the right of a hole, are
shown to exist in one dimension. While in two-dimensional case, four bound
states, with $s$-, $p_{x(y)}$- and $d_{x^{2}-y^{2}}$- symmetry respectively,
are found along the lines of $K_{x}=\pm K_{y}$. Away from the lines of $%
K_{x}=\pm K_{y}$, the degeneracy between $p_{x}$ and $p_{y}$ waves is
lifted. With decreasing the nearest-neighbor interaction $V$, the energies of the bound states increase and
merge into the particle-hole continuum gradually. The wave functions, the
dispersion relations and the existence regions of the bound states are
studied in details. For a given nearest-neighbor $V$, the number of bound states is different
in different regions of $\vec{K}$ space and a number distribution of bound
states is given for the nearest-neighbor $V=6t$ in two dimensions. The possible experimental observation
of the particle-hole bound states is also discussed.

Electron-hole bound state excitations (excitons) have been extensively
studied for many years and very recently, the excitonic Bose-Einstein
condensates have been realized experimentally in semiconductors \cite%
{butov,kasprzak}. We hope our study on the particle-hole bound states in
\textit{bosonic} systems would enrich our understanding of elementary
excitations in quantum many-body systems and stimulate more efforts on the
bound state phenomena in various systems such as magnets, superconductors
and atomic systems.

\noindent \textbf{Acknowledgements:} This work was supported by the NKBRSFC
under grants Nos. 2011CB921502, 2012CB821305, 2009CB930701, 2010CB922904,
NSFC under grants Nos. 10934010, 60978019 and NSFC-RGC under
grants Nos. 11061160490 and N-HKU748/10.


\begin{references}


\bibitem{molecule-review} For reviews, see L. D. Carr, D. DeMille, R. V. Krems and J. Ye, New J. Phys. \textbf{11}, 055049 (2009);
                          O. Dulieu and C. Gabbanini, Rep. Prog. Phys. \textbf{72}, 086401 (2009).
\bibitem{Jochim} S. Jochim, M. Bartenstein, A. Altmeyer, G. Hendl, S. Riedl,
C. Chin, J. H. Denschlag, R. Grimm, Science  \textbf{302}, 2101
(2003).
\bibitem{Greiner} M. Greiner, C. A. Regal and D. S. Jin, Nature  \textbf{426}, 537 (2003).
\bibitem{Volz} T. Volz, N. Syassen, D. M. Bauer, E. Hansis, S. D\"{u}rr and G. Rempe, Nature Physics \textbf{2}, 692 (2006).
\bibitem{Danzl1} J. G. Danzl, M. J. Mark, E. Haller, M. Gustavsson, R. Hart,
J. Aldegunde, J. M. Hutson and H.-C. N\"{a}gerl, Nature Physics
\textbf{6}, 265 (2010).
\bibitem{Ni} K. K. Ni, S. Ospelkaus, M. H. G. de Miranda, A. Pe'er, B. Neyenhuis, J. J. Zirbel,
S. Kotochigova, P. S. Julienne, D. S. Jin, J. Ye, Science
\textbf{322}, 231 (2008).
\bibitem{Ospelkaus} S. Ospelkaus, A. Pe'er, K. K. Ni, J. J. Zirbel, B. Neyenhuis, S. Kotochigova,
P. S. Julienne, J. Ye and D. S. Jin, Nature Physics \textbf{4}, 622
(2008).
\bibitem{Sage} J. M. Sage, S. Sainis, T. Bergeman and D. DeMille,  Phys. Rev. Lett. \textbf{94}, 203001 (2005).

\bibitem{Wang} D. Wang, J. T. Kim, C. Ashbaugh, E. E. Eyler, P. L. Gould and W. C. Stwalley, Phys. Rev. A \textbf{75},   032511
    (2007).
\bibitem{Sawyer} B. C. Sawyer, B. L. Lev, E. R. Hudson, B. K. Stuhl, M. Lara, J. L. Bohn and J. Ye,  Phys. Rev. Lett. \textbf{98},
    253002 (2007).
\bibitem{DeMille} D. DeMille, Phys. Rev. Lett.  \textbf{88}, 067901 (2002).
\bibitem{buchler} H. P. B\"{u}chler, E. Demler, M. Lukin, A. Micheli, N. Prokof'ev, G. Pupillo, and P. Zoller,  Phys. Rev. Lett.
    \textbf{98}, 060404 (2007).
\bibitem{Santos} L. Santos, G.V. Shlyapnikov, P. Zoller and M. Lewenstein,  Phys. Rev. Lett. \textbf{85}, 1791 (2000).

\bibitem{Cooper} N. R. Cooper, E. H. Rezayi and S. H. Simon,  Phys. Rev. Lett. \textbf{95}, 200402 (2005).
\bibitem{Baranov05} M. A. Baranov, Klaus Osterloh and M. Lewenstein,  Phys. Rev. Lett. \textbf{94}, 070404 (2005).
\bibitem{goral} K. G\'{o}ral, L. Santos and M. Lewenstein,  Phys. Rev. Lett. \textbf{88}, 170406 (2002).
\bibitem{Menotti} C. Menotti, C. Trefzger and M. Lewenstein,  Phys. Rev. Lett. \textbf{98}, 235301 (2007).
\bibitem{Lin} C. Lin, E. Zhao, and W. V. Liu, Phys. Rev. B \textbf{81}, 045115 (2010).
\bibitem{Bruder} C. Bruder, R. Fazio, and G. Sch\"{o}n, Phys. Rev. B \textbf{47}, 342 (1993).
\bibitem{Otterlo} A. van Otterlo, K. H. Wagenblast, R. Baltin, C. Bruder, R. Fazio and G. Sch\"{o}n, Phys. Rev. B \textbf{52}, 16176 (1995).
\bibitem{Niyaz} P. Niyaz, R. T. Scalettar, C. Y. Fong and G. G. Batrouni, Phys. Rev. B \textbf{50}, 362 (1994).
\bibitem{Sengupta} P. Sengupta, L. P. Pryadko, F. Alet, M. Troyer, and G. Schmid,  Phys. Rev. Lett. \textbf{94}, 207202 (2005).
\bibitem{Hassan} S. R. Hassan, L. de Medici, and A.-M. S. Tremblay, Phys. Rev. B \textbf{76}, 144420 (2007).
\bibitem{Iskin} M. Iskin and J. K. Freericks, Phys. Rev. A   \textbf{79}, 053634 (2009).
\bibitem{PAI} R. V. Pai and R. Pandit, Phys. Rev. B \textbf{71} 104508 (2005).
\bibitem{Chen} Y. C. Chen, R. G. Melko, S. Wessel, and Y. J. Kao, Phys. Rev. B \textbf{77}, 014524 (2008).

\bibitem{fisher} M. P. A. Fisher, P. B. Weichman, G. Grinstein, D. S. Fisher, Phys. Rev.  B \textbf{40}, 546 (1989).
\bibitem{stoof} D. van Oosten, P. vanderStraten, and H. T. C. Stoof, Phys. Rev.  A \textbf{63}, 053601 (2001).
\bibitem{Kovrizhin} D. L. Kovrizhin, G. V. Pai, and S. Sinha, Europhys. Lett.  \textbf{ 72}, 162 (2005).


\bibitem{fermion-exciton} F. B. Gallagher and S. Mazumdar, Phys. Rev. \textbf{B 56}, 15025 (1997);
                          W. Barford, Phys. Rev. \textbf{B 65}, 205118 (2002);
                          K. A. Al-Hassanieh, F. A. Reboredo, A. E. Feiguin, I. Gonz\'alez, and E. Dagotto,
                          Phys. Rev. Lett. \textbf{100}, 166403 (2008).


\bibitem{mattis} D. C. Mattis, The Theory of Magnetism, Vol. I: Statics and Dynamics, Springer-Verlag Series in Solid State Sciences
    (Berlin-New York, 1981).
\bibitem{Economou} E. N. Economou, Green's Functions in Quantum Physics, Third Edition, Springer-Verlag Series in Solid State Sciences
    (Berlin, 2006).

\bibitem{Nygaard} N. Nygaard, R. Piil, and K. M{\o}lmer, Phys. Rev.  A \textbf{78}, 023617 (2008).
\bibitem{Barford} W. Barford, Phys. Rev. \textbf{B 65}, 205118 (2002).

\bibitem{WORTIS} M. Wortis, Phys. Rev. \textbf{ 132}, 85 (1963).

\bibitem{Ye} J. W. Ye, J. M. Zhang, W. M. Liu, K. Y. Zhang, Y. Li, and W. P. Zhang, Phys. Rev.  A \textbf{83}, 051604(R) (2011).




\bibitem{kurn} D. M. Stamper-Kurn, A. P. Chikkatur, A. G\"{o}rlitz, S. Inouye, S. Gupta, D. E. Pritchard, and W. Ketterle, Phys. Rev. Lett. \textbf{83}, 2876 (1999).
\bibitem{Oosten} D. van Oosten, D. B. M. Dickerscheid, B. Farid, P. vanderStraten,  and H. T. C. Stoof, Phys. Rev. A \textbf{71},
    021601 (2005).
\bibitem{Rey} A. M. Rey, P. B. Blakie, G. Pupillo, C. J. Williams, and C. W. Clark, Phys. Rev. A \textbf{72}, 023407 (2005).
\bibitem{Ernst} P. T. Ernst, S. G\"{o}tze, J. S. Krauser, K. Pyka, D.-S. L\"{u}hmann,
D. Pfannkuche, and K. Sengstock, Nature Physics \textbf{6}, 56
(2010).
\bibitem{Bissbort}U. Bissbort, S. G\"{o}tze, Y. Li, J. Heinze, J. S. Krauser, M. Weinberg, C. Becker,
                 K. Sengstock, and W. Hofstetter, Phys. Rev. Lett. \textbf{106}, 205303
                 (2011).
\bibitem{Clement} D. Cl\'{e}ment, N. Fabbri, L. Fallani, C. Fort, and M. Inguscio,  Phys. Rev. Lett. \textbf{102}, 155301 (2009).

\bibitem{Bloch} I. Bloch, J. Dalibard, and W. Zwerger, Rev. Mod. Phys. {\textbf{80}}, 885 (2008).
\bibitem{Jaksch} D. Jaksch, C. Bruder, J. I. Cirac, C. W. Gardiner, and P. Zoller, Phys. Rev. Lett. {\textbf{81}}, 3108 (1998).
\bibitem{Mabrouk} N. Mabrouk and H. Berriche, J. Phys. B: At. Mol. Opt. Phys. {\textbf{41}}, 155101 (2008).
\bibitem{butov} L. V. Butov, C. W. Lai, A. L. Ivanov, A. C. Gossard, and D. S. Chemla, Nature \textbf{417}, 47 (2002).
\bibitem{kasprzak} J. Kasprzak, M. Richard, S. Kundermann, A. Baas, P. Jeambrun, J. M. J. Keeling, F. M. Marchetti,
M. H. Szyma\'{n}ska, R. Andr\'{e}, J. L. Staehli, V. Savona, P. B.
Littlewood, B. Deveaud, and L. S. Dang, Nature  \textbf{443}, 409 (2006).



\end{references}
\end{document}